# Integrating Metaverse Technologies in Medical Education: Examining Acceptance Factors Among Current and Future Healthcare Providers

Seckin Damar[1], Gulsah Hancerliogullari Koksalmis[2,3*]

[1] Department of Industrial Engineering, Faculty of Management, Istanbul Technical University, Istanbul, Turkey, damar17@itu.edu.tr, ORCID: 0000-0002-2713-2738

[2] Department of Industrial Engineering, Faculty of Management, Istanbul Technical University, Istanbul, Turkey, ghancerliogullari@itu.edu.tr, ORCID: 0000-0002-2551-541X

[3] Department of Industrial Engineering and Management Systems, University of Central Florida, Orlando, FL 32816, USA, gulsah.hancerliogullarikoksalmis@ucf.edu

[*] Corresponding author. E-mail address: ghancerliogullari@itu.edu.tr Telephone number: +90 530 312 48 91 (G. Hancerliogullari Koksalmis)

**Abstract**

This study investigates behavioral intention to use healthcare metaverse platforms among medical students and physicians in Turkey, where such technologies are in early stages of adoption. A multi-theoretical research model was developed by integrating constructs from the Innovation Diffusion Theory, Embodied Social Presence Theory, Interaction Equivalency Theorem and Technology Acceptance Model. Data from 718 participants were analyzed using partial least squares structural equation modeling. Results show that satisfaction, perceived usefulness, perceived ease of use, learner interactions, and technology readiness significantly enhance adoption, while technology anxiety and complexity have negative effects. Learner–learner and learner–teacher interactions strongly predict satisfaction, which subsequently increases behavioral intention. Perceived ease of use fully mediates the relationship between technology anxiety and perceived usefulness. However, technology anxiety does not significantly moderate the effects of perceived usefulness or ease of use on behavioral intention. The model explains 71.8% of the variance in behavioral intention, indicating strong explanatory power. The findings offer practical implications for educators, curriculum designers, and developers aiming to integrate metaverse platforms into healthcare training in digitally transitioning educational systems.

**Keywords** Healthcare metaverse . Medical education . Structural equation modeling . Partial least squares

## 1  Introduction

Within the field of medical education, continuous advances and the acceptance of new technologies have been fundamental. As the world progresses, so does the methodology of teaching and learning. A recent and noteworthy development in this field is the introduction of the metaverse.

The concept of "metaverse" originated with Neal Stephenson, who combined the terms "meta" and "universe" in his 1992 novel Snow Crash. Over the years, the metaverse has become synonymous with virtual, augmented, or mixed reality environments, where avatars and digital elements coexist. This fusion has been exemplified in gaming platforms like Second Life,

Minecraft, and Fortnite. However, beyond gaming, the metaverse has the ability to make radical changes in fields such as medical education by bringing unique learning features (Sandrone, 2022).

Educational techniques have continuously evolved, transitioning from traditional methods to technology-based approaches. The Covid-19 pandemic in 2020, however, underscored the necessity of remote learning and triggered a significant rise in interest and investment in metaverse technologies during that period. Today, college students can access various media innovations, including "virtual reality" (VR), "augmented reality" (AR), and "mixed reality" (MR). They can use these tools to participate in online video, network education, and broadcasts. Additionally, students can experience VR, 3D visuals, and AR. This integration creates a parallel virtual world alongside the real one. It connects the two realms and establishes novel human-computer interaction modes. Furthermore, this expansion of sensory experiences enhances physical immersion. These advancements not only alter students' lives but also reshape their cognitive habits, lifestyles, and behavioral patterns significantly (Ge, 2022).

In this study, particular emphasis is placed on the implementation of metaverse technologies in medical education, specifically targeting healthcare providers—both current practitioners (physicians) and future healthcare professionals (medical students). The rationale behind this approach is rooted in the capabilities of metaverse technologies to transform educational environments by providing immersive, simulation-based learning experiences. These technologies are expected to close the gap between theoretical knowledge and practical application, thereby preparing medical students to transition seamlessly into clinical roles as healthcare providers.

The emergence of the metaverse has significantly transformed medical education, particularly in anatomy and surgical training. At Case Western Reserve University School of Medicine, the GARLA (Gross Anatomy, Radiology, Living Anatomy) program integrates anatomy, radiology, and physical examination skills through virtual labs. Using HoloAnatomy, students efficiently learn human structure and function while interpreting radiology images and practicing on standardized patients, bridging classroom learning with clinical practice (Wish-Baratz et al., 2020). Similarly, the National University of Singapore Yong Loo Lin School of Medicine collaborates with Microsoft on Project Polaris, incorporating HoloLens 2 to enhance anatomy education through mixed reality experiences (Microsoft-News-Center, 2022). In South Korea, a metaverse-based surgical training program equips residents with head-mounted virtual reality devices, providing live 360-degree video recordings from an operating room at Seoul National University Bundang Hospital (Koo, 2021). These immersive technologies also advance orthopedic surgery training, including joint replacements and fracture repair, offering realistic, data-driven simulations that enhance surgical skills (Bansal et al., 2022).

To clarify, the term "healthcare metaverse" in this study refers to an interactive, immersive, and simulation-based virtual environment designed for medical education and training. It encompasses VR, AR, and MR technologies that enable medical students and healthcare professionals to develop clinical skills through realistic, scenario-based learning experiences. Unlike healthcare metaverse applications focused on telemedicine or patient care, this study defines it as a technology-enhanced educational ecosystem where users can engage in virtual simulations, interactive case studies, and skill-based assessments. By providing real-time interaction, experiential learning, and accessibility beyond physical training environments, the healthcare metaverse serves as a transformative tool for bridging the gap between theoretical knowledge and practical application in medical education.

The participants in this study include both medical students and practicing physicians, representing the future and current healthcare workforce. Their behavioral intentions to adopt metaverse technologies were measured, as the technology has not yet been fully implemented in their educational or clinical environments. By focusing on behavioral intentions rather than actual usage, this study aims to capture their openness to adopting these technologies in the future, providing insights into how such innovations may be integrated into educational and clinical settings. Importantly, medical students are also potential future educators, who may contribute to the dissemination of metaverse technologies within medical education.

In this study, a bibliometric analysis was performed to examine the existing body of literature on the acceptance of healthcare metaverse technologies in medical education, using VOSviewer software as a key tool for identifying patterns and relationships within the research. A detailed bibliometric analysis of the healthcare metaverse is available in Damar and Koksalmis (2024). A comprehensive query was applied: '(TITLE-ABS-KEY ("metaverse" OR "healthcare metaverse") AND TITLE-ABS-KEY ("structural equation model*" OR "sem" OR "technology acceptance model*" OR "tam" OR "intent*" OR "accept*" OR "adopt*") AND TITLE-ABS-KEY ("health*" OR "medical" OR "medicine") AND TITLE-ABS-KEY ("educat*")) AND (EXCLUDE (DOCTYPE, "re") OR EXCLUDE (DOCTYPE, "cr"))'. As of October 2024, this search yielded 70 relevant publications. Although efforts were made to exclude review articles, it was found that some studies still contained review elements, which necessitated further refinement. After a detailed review, it was determined that only 11 studies fully met the inclusion criteria. This limited number of studies has been included in the literature review section to ensure a focused and thorough analysis of existing research.

While prior research has explored the adoption of digital learning technologies in medical education, many studies have relied on traditional acceptance models, such as the Technology Acceptance Model and Unified Theory of Acceptance and Use of Technology. These studies emphasize key determinants like perceived usefulness (PU) and perceived ease of use (PEOU), yet they often overlook other critical adoption factors such as satisfaction, presence, trialability, complexity, compatibility, and relative advantage. Additionally, previous research has frequently focused on small sample sizes or single user groups, limiting generalizability. This study addresses these gaps by integrating Innovation Diffusion Theory to assess the diffusion of metaverse technologies, Extended Spatial Presence Theory to examine learner engagement in virtual spaces, and Interaction Equivalency Theorem to analyze the role of interaction types in enhancing educational outcomes. By incorporating these perspectives, this research provides a more comprehensive understanding of metaverse adoption in medical education and offers a broader theoretical framework compared to previous studies. Furthermore, previous research has predominantly focused on medical students, overlooking the perspectives of practicing physicians who play a crucial role in the implementation and future dissemination of these technologies. Since medical students will transition into clinical roles, and experienced physicians often contribute to medical training, understanding the attitudes of both current and future healthcare professionals is necessary for developing effective adoption strategies. This study addresses this gap by including both groups, offering a more holistic assessment of metaverse acceptance in medical education. Another limitation of prior studies is their narrow focus on direct adoption factors without considering how psychological variables influence acceptance. This study uniquely examines the moderation effect of technology anxiety and the mediation role of PEOU between technology anxiety and perceived usefulness. By doing so, it provides deeper insights into how user anxiety indirectly influences adoption decisions, emphasizing the importance of reducing technological complexity and enhancing usability to facilitate acceptance.

The primary objective of this study is to examine the adoption of healthcare metaverse technologies in medical education by assessing the factors that influence behavioral intention. While prior studies have emphasized factors such as usefulness and ease of use, fewer have examined the psychological barriers, presence, and adoption dynamics through multiple theoretical lenses. Therefore, to address these gaps, we propose the following research questions (RQs):

RQ1 How do key innovation attributes (relative advantage, compatibility, complexity, trialability) affect the adoption of healthcare metaverse technologies?

RQ2 How does the sense of presence in the healthcare metaverse impact PU and behavioral intention?

RQ3 How do learner-teacher and learner-learner interactions influence satisfaction, and how does satisfaction subsequently impact behavioral intention to use the healthcare metaverse?

RQ4 How do PU and PEOU influence behavioral intention to use healthcare metaverse technologies in medical education?

RQ5 To what extent does technology readiness influence PEOU and presence in the healthcare metaverse?

RQ6 How does technology anxiety influence PU and PEOU, and how does it moderate the relationships between PU, PEOU, and behavioral intention?

RQ7 Does PEOU mediate the relationship between technology anxiety and PU?

This research makes a significant contribution to the field in several ways. First, it expands the theoretical landscape by integrating multiple models, offering a more nuanced and multidimensional framework for understanding metaverse adoption in medical education. Second, by incorporating medical students and practicing physicians, it provides a more representative analysis of key user groups that will shape the future of metaverse-based education. Third, it introduces novel insights into the psychological and contextual barriers to adoption, particularly focusing on the impact of technology anxiety and ease of use. Finally, as one of the first studies of its kind in Turkey, this research adds valuable empirical evidence to a growing field, providing actionable insights for educators, policymakers, and developers seeking to integrate metaverse technologies into medical curricula. The findings extend beyond medical education, contributing to a broader understanding of immersive technology adoption in professional training environments.

The rest of the paper is structured as follows: In Section 2, the study model and hypotheses are developed, the model is examined, and a comprehensive literature review is provided with an emphasis on the theoretical foundations of technology acceptance. Data collection procedure, measures taken, and data analysis strategy are described in depth in Section 3 of the study's methodology. Then, a comprehensive description of the findings of the analysis is provided in Section 4. Section 5 continues with a discussion and conclusion, acknowledging the observed limitations and suggesting future directions of work. The implications of the findings, both theoretical and practical, are also discussed.

## 2  Theoretical background and hypotheses development

There are numerous well-known models that have developed throughout time in the literature on acceptance theory. Fishbein and Ajzen (1975) suggested the "Theory of Reasoned Action" (TRA), which emphasized personal attitudes and subjective norms in technology acceptance. Ajzen (1985) further extended this with the "Theory of Planned Behavior" (TPB), incorporating the concept of perceived behavioral control. "Innovation-Diffusion Theory" (IDT) by Rogers (1983) categorizes individuals according to their acceptance behavior of new technologies and emphasized innovation characteristics. Davis (1989) introduced the "Technology Acceptance Model" (TAM), which highlights that a person's behavioral intention to accept new technologies is driven by perceived usefulness and PEOU. Venkatesh et al. (2003) developed the "Unified Theory of Acceptance and Use of Technology" (UTAUT), which combined performance expectancy, effort expectancy, and social influence. In UTAUT2, Venkatesh et al. (2012) added price value and pleasure-driven motivation as additional factors, all of which were moderated by demographic and experiential variables. Additionally, the "Embodied Social Presence Theory" by Mennecke et al. (2010) explains how users develop a sense of presence in virtual environments through avatar-based interactions, while the "Interaction Equivalency Theorem" by Anderson (2003) highlights the role of learner-instructor, learner-content, and learner-learner interactions in shaping effective learning experiences. Collectively, these models provide frameworks for understanding the complex processes associated with the acceptance and use of new technology.

## 2.1 Innovation-diffusion theory

The IDT, proposed by Rogers (1983), is among the earliest frameworks that examine the factors influencing an individual's decision to adopt new technologies. The theory suggests that innovation adoption is driven by the need to reduce uncertainty. To achieve this, individuals seek and process information about the technology, forming beliefs about its use. These beliefs ultimately shape their decision to either accept or reject the innovation (Nor et al., 2010).

IDT has primarily focused on how the perceived characteristics of innovations and the innovativeness of adopting organizations affect adoption rates. Rogers (2003) highlights that relative advantage, compatibility with existing workflows, complexity, trialability, and observability shape how quickly new technologies are adopted. When innovations are seen as beneficial, easy to integrate, and observable within an industry, adoption tends to accelerate (García-Avilés, 2020).

## 2.2 Interaction equivalency theorem

The Interaction Equivalency Theorem (EQuiv), introduced by Anderson (2003), builds on the idea that effective learning can be achieved when at least one of the three key interactions, namely learner-content, learner-instructor, or learner-learner, is strong. If one of these interactions is at a high level, the other two can be minimal or even absent without negatively impacting the learning experience. While incorporating multiple high-level interactions can enhance satisfaction, it may also increase costs and time requirements (Song, 2010).

The "Three Types of Interaction" model, originally proposed by Moore (1989), was the first systematic framework to define interaction as a critical component of distance education. As an extension of this model, the EQuiv framework was developed to provide a theoretical basis for evaluating the necessary balance of different interaction types in educational settings (Miyazoe & Anderson, 2013).

## 2.3 Embodied social presence theory

Mennecke et al. (2010) proposed the Embodied Social Presence Theory (ESPT), which explains how users develop a sense of presence in virtual environments through self-recognition, interaction, and engagement. Users first recognize their own avatars, then interact with others, creating a shared sense of presence. This process involves observing avatars, engaging in activities, and interpreting virtual interactions similarly to real-world communication. Ultimately, ESP is a dynamic cycle, where recognition and interaction continuously shape the perception of presence (Mennecke et al., 2010).

Mennecke et al. (2011) describe embodied presence as a multidimensional construct influenced by technical preparation, content variables, and user variables. Technical preparation encompasses interface design, interaction quality, and sensory stimulation, which help users feel mentally and emotionally immersed in the virtual environment. Content variables, such as avatar appearance, movement, and environmental elements, contribute to a sense of realism, allowing users to perceive similarities between virtual and physical spaces. User variables, including prior experience, cognitive state, and self-efficacy, further shape the extent to which individuals feel present in the virtual world. These components align with the broader framework of ESPT, which suggests that a strong sense of presence emerges when individuals engage meaningfully with virtual stimuli, reinforcing their identity and interaction within the digital environment (Mennecke et al., 2011).

## 2.4 Technology acceptance model

TAM stands out as one of the few models incorporating psychological elements influencing technology acceptance. Derived from the TRA model, TAM characterizes and anticipates individuals' behaviors within particular contexts (Sarlan et al., 2013). It stands out as a highly robust theory frequently employed by scholars to examine how individuals accept technology (Pandya & Kumar, 2023). The concept originates from an initial effort to understand what influences users to accept computer systems. Davis (1989) suggests that a user's acceptance behavior depends on their intention to use the systems, influenced by their attitude. This perspective is mainly shaped by two factors: PU and PEOU (Abroud et al., 2015).

## 2.5 Previous studies

The reviewed studies on healthcare metaverse acceptance in medical education reveal a diverse application of TAM, UTAUT2, and advanced analytical techniques such as SEM and machine learning to assess user adoption. These studies emphasize key determinants like PU, PEOU, perceived value, social influence, and hedonic motivation, while also highlighting emerging factors such as trialability, observability, and personal innovativeness. Table 1 also summarizes these studies.

Several studies employed TAM to assess behavioral intentions toward metaverse-based learning in medical education. Yap et al. (2024) explored an AI-powered immersive platform combining generative AI and art therapy to enhance healthcare students' understanding of patient experiences. The study found that perceived playfulness, compatibility, and social norms significantly influenced behavioral intentions, though the small sample of 16 participants limited generalizability. Nagadeepa et al. (2023) investigated 214 healthcare students' perceptions of metaverse-based learning and found that PU, PEOU, system usability, and social influence were key predictors, with 48% of participants being early adopters. However, a gap was identified between students' awareness and actual experience with metaverse platforms. Alawadhi et al. (2022) examined 435 medical students, incorporating personal innovativeness and perceived enjoyment alongside core TAM components. Their findings confirmed that these factors played a significant role in adoption, though the study's reliance on social media for data collection may have introduced biases.

Some studies extended the TAM framework using UTAUT2, offering a broader perspective on metaverse adoption. Kalınkara and Özdemir (2024) applied UTAUT2 to analyze midwifery students' acceptance of metaverse technology for anatomy education, using PLS-SEM on 47 students' responses. The study identified social influence, facilitating conditions, and habit as significant predictors, explaining 75.3% of the variance in behavioral intention. Abdulmuhsin et al. (2024) focused on 278 medical educators, adapting UTAUT2 to examine knowledge management-driven metaverse adoption. Their findings highlighted perceived value and hedonic motivation as critical factors, with interaction indirectly influencing behavioral intention through knowledge-sharing processes.

Several studies combined TAM with SEM and ML techniques to enhance predictive accuracy and explore adoption patterns in greater depth. Almarzouqi et al. (2024) assessed 89 dental students' adoption of metaverse technologies, focusing on perceived value and satisfaction as key determinants. Salloum et al. (2024) conducted a large-scale study on 833 students, integrating artificial neural networks (ANN) with SEM to examine metaverse adoption in dental education. Their findings revealed that mobility and accessibility were the strongest predictors of user adoption. Aburayya et al. (2023) and Salloum et al. (2023) employed hybrid SEM-ML models to explore TAM constructs alongside additional factors such as trialability, ubiquity, observability, and perceived enjoyment. AlHamad et al. (2022), using a hybrid SEM-ML approach on 477 medical students, confirmed the importance of perceived value, although the study was limited by its focus on only PU and PEOU. Similarly, Almarzouqi et al. (2022b) conducted a large-scale study with 1858 students, finding that user satisfaction played a crucial role in adoption, with ML techniques providing deeper insights into how trialability, observability, and personal innovativeness influenced metaverse acceptance.

These findings reinforce that TAM remains a foundational framework for understanding metaverse adoption in healthcare education but benefits from integration with UTAUT2, interaction theories, and perceived value-based models to incorporate additional adoption factors. While PU and PEOU are central predictors, variables such as social norms, perceived enjoyment, perceived value, trialability, and observability have emerged as significant in immersive learning environments. However, many studies have relied on limited theoretical models or small sample sizes, affecting the generalizability of their findings. To address gaps in previous research, this study adopts a multi-theoretical approach, examining the behavioral intentions of both medical students and practicing physicians. Since students will transition into clinical roles, assessing both groups together offers a holistic view of metaverse adoption in education and practice. This broader framework enables a deeper and more nuanced understanding of its implications for medical training.

## 2.6 Hypotheses development

A comprehensive framework has been developed by integrating constructs from multiple theoretical models to provide a deeper understanding of healthcare metaverse adoption. IDT has been incorporated to capture key factors influencing the adoption of innovative technologies, and RQ1 is attempted to be solved here by examining how attributes such as compatibility, relative advantage, complexity, and trialability influence adoption. Compatibility has been included to assess the degree to which metaverse technologies align with users' existing professional values, workflows, and technological infrastructure. The likelihood of adoption increases when an innovation seamlessly integrates with users' existing experiences, values, and needs, as it reduces uncertainty and minimizes disruption (Moore & Benbasat, 1991; Sahin, 2006; Tornatzky & Klein, 1982). In structured environments such as healthcare, ensuring that new systems align with professional workflows is essential for successful diffusion. Relative advantage has been added to examine whether these technologies offer measurable improvements over conventional clinical training and medical education methods, as perceived benefits strongly influence adoption decisions. Within the framework of IDT, relative advantage is recognized as one of the most influential predictors of an innovation's adoption rate, as users are more likely to embrace technologies that provide clear improvements over existing solutions (Sahin, 2006). When healthcare metaverse applications offer enhanced accessibility, interactivity, or training effectiveness, their perceived advantage over traditional learning methods strengthens adoption. Complexity has been considered to account for potential usability challenges, recognizing that the perceived difficulty of using new technologies can affect acceptance. Technologies that are easier to understand and use are more likely to be adopted, as excessive complexity can deter potential users from engaging with them (Moore & Benbasat, 1991; Tornatzky & Klein, 1982). Given the immersive and technically sophisticated nature of metaverse-based platforms, it is crucial to ensure that their interfaces and functionalities are intuitive to encourage engagement among healthcare professionals. Additionally, trialability has been incorporated to highlight the significance of allowing users to test the technology before full integration, which is particularly relevant in high-stakes fields like medicine, where risk mitigation is essential. When users have opportunities to experiment with an innovation in a controlled setting, they are more likely to develop confidence in its use, leading to a faster and smoother adoption process (Akour et al., 2022). By providing pilot programs or structured training sessions, healthcare institutions can support a more seamless transition to metaverse-based training environments.

To further enhance the understanding of engagement with virtual environments, ESPT has been integrated, and RQ2 is attempted to be solved here by exploring how presence influences PU and behavioral intention. Additionally, imagination is examined under ESPT as a key cognitive process that enables individuals to engage more deeply with virtual environments. By allowing users to conceptualize virtual interactions and anticipate the benefits of metaverse technologies, imagination enhances both presence and PU. Embodied presence has been included to examine how users perceive immersion in the virtual environment, which is critical for effective learning and interaction in metaverse-based applications. The emergence of a sense of presence can influence psychological and physiological aspects such as role involvement, enjoyment, and the development of prosocial relationships (Zhang et al., 2022). Since presence fosters deeper engagement, improving these aspects in metaverse-based education can enhance user experiences and learning outcomes. Imagination has been identified as a key user variable that influences PU and presence by enabling individuals to conceptualize virtual interactions and anticipate the potential benefits of metaverse technologies in clinical and educational contexts.

The ability to visualize scenarios in virtual environments enhances problem-solving skills and creativity, particularly in open-ended learning situations (Huang et al., 2010). Creative visualization techniques can support learners in immersing themselves in metaverse-based simulations, making training more effective and engaging.

Furthermore, TAM has been integrated to assess users' PU and PEOU, which are critical determinants of technology adoption, and RQ4 is attempted to be solved here by analyzing their impact on behavioral intention. In metaverse-based medical education, PU is expected to influence adoption, as healthcare professionals and students are more likely to embrace the technology if they perceive it as beneficial for clinical training and skill development. Given that healthcare professionals often operate under time constraints, ensuring that metaverse platforms are user-friendly and intuitive will be key to promoting acceptance.

Technology readiness has been considered to assess users' preparedness and willingness to adopt emerging technologies, and RQ5 is attempted to be solved here by investigating its effect on presence and PEOU. Technology readiness is recognized as a crucial factor influencing the likelihood of adoption, as it determines whether individuals perceive new systems as enablers or barriers to achieving their goals (Almaiah, Alfaisal, Salloum, Al-Otaibi, et al., 2022; Parasuraman, 2000). Individuals with higher technology readiness are expected to experience greater presence in virtual environments, as they are more comfortable engaging with immersive technologies. Additionally, technology readiness influences PEOU, as individuals with higher readiness are likely to navigate the healthcare metaverse more efficiently, reducing cognitive load and increasing overall engagement.

While much of the existing research on technology adoption in medical education has focused on usefulness and ease of use, psychological barriers, such as technology anxiety, and their moderating and mediating effects have received less attention. Technology anxiety has been introduced as both a direct predictor of PEOU and PU and a moderating factor in the adoption process. RQ6 is attempted to be solved here by investigating how technology anxiety influences PU and PEOU and how it moderates the relationships between PU, PEOU, and behavioral intention. Individuals with higher anxiety toward technology are generally less inclined to adopt new systems, even if they recognize potential benefits (Bhatt, 2022). Anxiety may cause users to find metaverse platforms more difficult to navigate, reducing their ease of use perceptions and potentially limiting their ability to see these technologies as beneficial. Additionally, technology anxiety may weaken the influence of PU and PEOU on behavioral intention, preventing adoption even when users acknowledge the potential value of the system.

This study also attempts to solve RQ7 here by analyzing the mediation effect of PEOU between technology anxiety and PU. Since ease of use directly influences PU, understanding how technology anxiety affects this relationship is crucial in explaining indirect psychological influences on adoption behavior. Users experiencing higher anxiety may struggle to interact with metaverse-based medical education platforms, leading to lower PEOU, which in turn reduces their perception of the system's usefulness.

Finally, Interaction Equivalency Theorem has been applied by incorporating learner-teacher interaction and learner-learner interaction, and RQ3 is attempted to be solved here by examining their role in satisfaction and behavioral intention in metaverse-based medical education. High levels of interaction between learners and instructors, as well as among peers, are associated with more satisfying educational experiences, as they promote engagement, knowledge exchange, and collaboration (Miyazoe & Anderson, 2013). Since higher satisfaction leads to increased adoption intention, satisfaction serves as a key intermediary factor, linking interaction quality to behavioral intention. A well-designed virtual learning environment should facilitate meaningful interactions to maximize educational effectiveness and user satisfaction, which in turn may enhance adoption likelihood.

### 2.6.1 Behavioral intention

Behavioral intention to use (BIU) is "the degree to which a person has formulated conscious plans to perform or not perform some specified future behavior" (Warshaw & Davis, 1985). It reflects the possibility of users' attitudes and beliefs turning into real behavior. This concept refers to the desire to use a technology and has been extensively researched in the TAM literature. In general, behavioral intention depends on individuals' positive or negative perceptions and ultimately influences their decision to interact with a product or service (Jeong & Kim, 2023). In this study, BIU pertains to the willingness of medical professionals and students to employ the healthcare metaverse technology for medical education purposes. Given the relatively limited number of individuals actively engaging with the health metaverse in education, the research prioritized understanding the factors influencing behavioral intention within the system. The study identifies several constructs that influence BIU, including PEOU, PU, perceived trialability, technology readiness, imagination, and technology anxiety. The model presented in the research, visually represented in Figure 1, illustrates the relationships among these constructs.

**Table 1** Summary of healthcare metaverse studies in medical education

| Authors | Methods/Model used | Constructs | Number of participants | Key findings |
| --- | --- | --- | --- | --- |
| Yap et al. (2024) | TAM | Perceived playfulness, perceived compatibility, social norms | 16 | Small sample size limits generalizability but highlights potential for empathy development. |
| Nagadeepa et al. (2023) | TAM | PU, PEOU, system usability, social influence | 214 | PU and PEOU are key; gap between awareness and actual experience. |
| Alawadhi et al. (2022) | TAM | Personal innovativeness, perceived enjoyment, perceived usefulness | 435 | Personal innovativeness and enjoyment significantly influence intention; social media may introduce biases. |
| Kalınkara and Özdemir (2024) | UTAUT2 | Social influence, facilitating conditions, habit | 47 | Social influence, facilitating conditions, and habit are critical predictors of intention. |
| Abdulmuhsin et al. (2024) | UTAUT2 | Perceived value, hedonic motivation, interaction | 278 | Perceived value and hedonic motivation are critical, with knowledge sharing as an indirect influence. |
| Almarzouqi et al. (2024) | PLS-SEM and ML | Perceived value, perceived satisfaction | 89 | Satisfaction and perceived value are critical factors in adoption. |

**Table 1** (continued)

| Authors | Methods/Model used | Constructs | Number of participants | Key findings |
| --- | --- | --- | --- | --- |
| Salloum et al. (2024) | SEM and ANN | Mobility, accessibility | 833 | Mobility and accessibility are key; ANN provides higher prediction accuracy. |
| Aburayya et al. (2023) | PLS-SEM and ML | Perceived value, ubiquity | 369 | Perceived value significantly affects intention; ML gives deeper insights. |
| Salloum et al. (2023) | SEM and ML | Trialability, observability, enjoyment, ubiquity, perceived value | 879 | Trialability and enjoyment are key predictors; methodology provides solid foundation. |
| AlHamad et al. (2022) | SEM and ML | Perceived value, perceived enjoyment | 477 | Perceived value and enjoyment are key; limited focus on TAM constructs. |
| Almarzouqi et al. (2022b) | SEM and ML | User satisfaction, personal innovativeness, perceived trialability, observability | 1858 | User satisfaction and personal innovativeness are key; advanced ML techniques provide nuanced insights. |

## 2.6.2 Satisfaction

Satisfaction (SAT) is defined as "the summary psychological state resulting when the emotion surrounding disconfirmed expectations is coupled with the consumer's prior feelings about the consumption experience" (Oliver, 1981). In healthcare metaverse adoption for medical education, satisfaction reflects how well the technology meets the expectations of students and professors. When users have positive experiences such as intuitive navigation, immersive learning and effective instruction, they are more likely to be satisfied with the system. Satisfaction, in turn, increases their desire to continue using metaverse for educational purposes.

Empirical studies highlight the strong relationship between satisfaction and behavioral intention in educational and healthcare technology adoption. Research confirms that higher satisfaction levels encourage continued use and engagement with digital learning environments (Di Natale et al., 2024; Han & Sa, 2022; Rajeh et al., 2021).

H1. "Satisfaction positively influences behavioral intention to use healthcare metaverse in medical education."

## 2.6.3 Perceived usefulness

PU is "the degree to which an individual believes that using a particular system would enhance his or her job performance" (Davis, 1989, p. 320). According to the TAM, PU is a key determinant of BIU, as users are more likely to adopt a system they perceive as beneficial. In the healthcare metaverse, PU influences BIU because medical professionals and students are more inclined to engage with the system if it enhances clinical training, decision-making, and professional development. A metaverse-based learning environment that offers realistic simulations, hands-on training, and interactive case studies will be perceived as a valuable educational tool, increasing adoption.

Beyond BIU, PU also affects user satisfaction, which depends on how well a system meets expectations. If the healthcare metaverse delivers engaging, efficient, and effective learning experiences, users will feel satisfied. However, if the perceived benefits do not match reality, satisfaction may decline, reducing long-term use.

Studies confirm that PU positively predicts BIU, as users prefer technologies they find beneficial (Akour et al., 2022; İbili et al., 2023; Toraman, 2022; Wu & Yu, 2023). Al-Adwan et al. (2023) emphasized that demonstrating the metaverse's advantages over traditional learning is key to increasing PU. Also, Alfaisal et al. (2024) found a significant positive impact of PU on students' metaverse adoption. PU is also strongly linked to satisfaction. Users who perceive a system as useful are more likely to have positive learning experiences, leading to higher satisfaction levels. Empirical research also supports this connection (AL-Hawamleh, 2024; Del Barrio-García et al., 2015; Hancerliogullari Koksalmis & Damar, 2022; Legramante et al., 2023). Consequently, the hypothesis can be stated as follows:

H2. "Perceived usefulness positively influences the behavioral intention to use healthcare metaverse in medical education."

H3. "Perceived usefulness positively influences satisfaction with the healthcare metaverse in medical education.

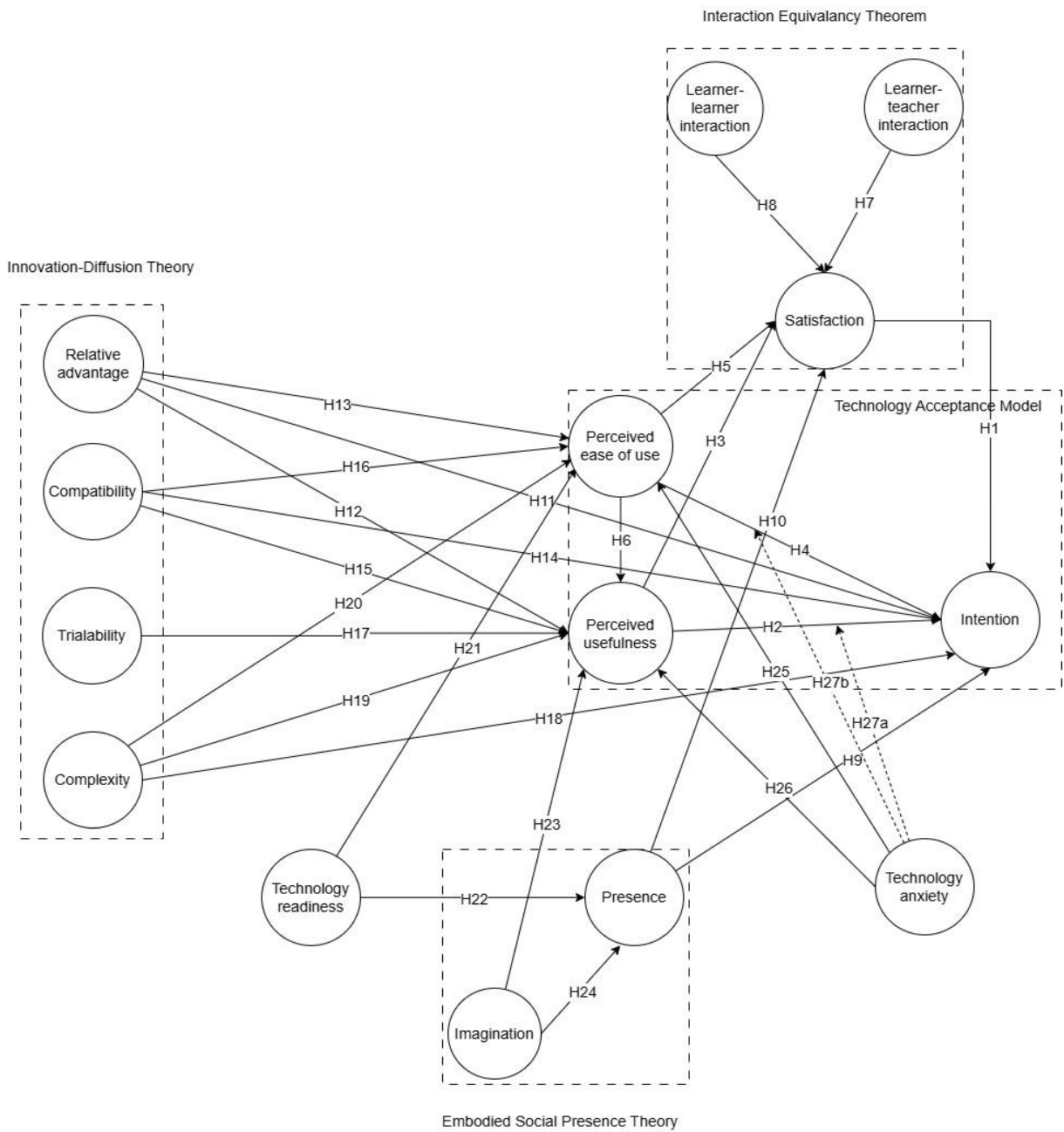

**Fig. 1** Research model

## 2.6.4 Perceived ease of use

PEOU is "the degree to which a person believes that using a particular system would be free of effort" (Davis, 1989, p. 320). In the context of the healthcare metaverse, PEOU reflects how easily medical professionals and students can interact with the system, perform tasks, and integrate the technology into their learning or teaching environments.

A system that is intuitive, well-structured, and user-friendly minimizes cognitive overload and reduces frustration, making it more likely to be adopted. According to the TAM, PEOU influences both PU and BIU. Users who find a system simple to use are more inclined to view it as advantageous and show a greater willingness to adopt it. Additionally, a positive user experience in terms of ease of use contributes to satisfaction, as users who can efficiently operate the system are more likely to have enjoyable and productive interactions, reinforcing continued use.

Prior research validates PEOU's role in technology adoption. Studies confirm that PEOU has a direct positive impact on BIU, as users prefer systems that are simple and effortless to use (Alawadhi et al., 2022; Alfaisal et al., 2022; Almarzouqi et al., 2022b; Mostafa, 2022). Moreover, empirical evidence demonstrates a strong relationship between PEOU and PU, highlighting that an easy-to-use system is more likely to be perceived as useful (Akour et al., 2022; Alawadhi et al., 2022; Damar & Koksalmis, 2023; Toraman, 2022). Additionally, research in digital learning environments establishes a positive relationship between PEOU and satisfaction, indicating that a seamless user experience enhances overall satisfaction with the system (AL-Hawamleh, 2024; Almulla, 2024; Alqahtani et al., 2022; Hancerliogullari Koksalmis & Damar, 2022). Based on this, the following hypotheses could be developed:

H4. "Perceived ease of use positively influences the intention to use healthcare metaverse in medical education."

H5. "Perceived ease of use positively influences satisfaction with the healthcare metaverse in medical education."

H6. "Perceived ease of use positively influences the perceived usefulness of healthcare metaverse in medical education."

## 2.6.5 Learner- learner and learner-teacher interaction

Interaction is a key component of effective learning experiences, particularly in digital and immersive environments. The Interaction Equivalency Theorem suggests that high-quality interaction between learners and instructors, or among learners themselves, significantly enhances engagement, learning outcomes, and satisfaction.

Learner-teacher interaction (LTI) refers to the communication that occurs "between the learner and the expert who prepared the subject material, or some other expert acting as an instructor" (Moore, 1989). A strong teacher presence in the metaverse can enhance motivation and comprehension by addressing learners' concerns and providing personalized support. On the other hand, learner-learner interaction (LLI) involves engagement "between one learner and other learners, alone or in group settings, with or without the real-time presence of an instructor" (Moore, 1989). In a healthcare metaverse environment, students working together in simulated clinical scenarios or engaging in group discussions can improve their understanding of medical concepts and foster teamwork skills.

When both learner-teacher and learner-learner interactions are present and effective, users are more likely to feel satisfied with their learning experiences. Earlier studies have shown that learner-teacher interaction and learner-learner interaction are key factors influencing satisfaction (Hancerliogullari Koksalmis & Damar, 2022; Hettiarachchi et al., 2021; Nagy, 2018; Singh et al., 2024). Studies highlight that learner-teacher interaction enhances engagement and overall learning experiences by providing structured guidance and immediate feedback. Similarly, learner-learner interaction fosters peer collaboration and active engagement, leading to higher satisfaction with digital learning environments.

H7. "Learner-teacher interaction positively influences satisfaction with the healthcare metaverse in medical education."

H8. "Learner-learner interaction positively influences satisfaction with the healthcare metaverse in medical education."

### 2.6.6 Presence

The concept of presence (PRE) encompasses meanings such as existence, being, or reality. In a virtual environment, it refers to the sensation of experiencing a computer-mediated virtual space controlled by a user who is not physically present. It also describes the perception of being physically close to a remotely located individual, despite the actual physical distance (Oh & Yoon, 2014). According to ESPT, stronger presence fosters natural interactions, increasing cognitive and emotional involvement. In the healthcare metaverse, presence allows medical students to experience realistic training simulations, reinforcing learning and skill acquisition.

A strong sense of presence positively influences BIU as users are more likely to adopt a system that feels immersive and realistic. When virtual medical environments replicate real-world scenarios effectively, users gain confidence in the system's value, increasing adoption. Presence also contributes to satisfaction by improving engagement and reducing frustration. A highly immersive environment enhances the user experience, making the learning process more enjoyable and effective, whereas a lack of presence may lead to lower satisfaction and decreased adoption.

Studies confirm that higher presence levels increase BIU by making virtual experiences more engaging and meaningful (Zhang et al., 2022). Additionally, research in virtual learning environments shows that presence enhances satisfaction by improving immersion and perceived realism (Kim & Zhang, 2011).

H9: "Presence positively influences behavioral intention to use the healthcare metaverse in medical education."

H10: "Presence positively influences satisfaction with the healthcare metaverse in medical education."

### 2.6.7 Relative advantage

Relative advantage (RA) is "the degree to which an innovation is perceived as being better than the idea it supersedes" (Rogers, 2003). According to IDT, relative advantage is a key driver of technology adoption, as individuals are more likely to adopt innovations that offer significant improvements over existing solutions. In the context of healthcare metaverse adoption, relative advantage refers to the extent to which the metaverse enhances learning outcomes, clinical training, and interactivity compared to traditional methods.

A greater perceived relative advantage positively influences PU, as users are more likely to adopt a system that offers clear benefits over conventional methods. If medical students and professionals perceive that the metaverse improves their learning efficiency, procedural skills, and engagement, they will find the technology more useful. Relative advantage also affects PEOU by shaping user expectations. If a technology is seen as superior and transformative, users may be more willing to invest effort in learning it, perceiving it as intuitive and worthwhile. Additionally, higher relative advantage strengthens BIU, as individuals prefer adopting innovations that significantly enhance their performance or experience.

Empirical research confirms the strong impact of relative advantage on technology adoption. Studies demonstrate that perceived relative advantage positively influences BIU, as users are more likely to adopt a technology they see as superior to existing alternatives (Ayanwale & Molefi, 2024; Lee et al., 2011; Raman et al., 2021). Additionally, research shows that relative advantage enhances both PU and PEOU, as users who perceive clear benefits from a system are more likely to find it useful and easy to use (Al-Rahmi et al., 2021; Al-Rahmi et al., 2019; Günay & Toraman, 2023; Lee et al., 2011).

H11: "Relative advantage positively influences behavioral intention to use the healthcare metaverse in medical education."

H12: "Relative advantage positively influences perceived usefulness of the healthcare metaverse in medical education."

H13: "Relative advantage positively influences perceived ease of use of the healthcare metaverse in medical education."

### 2.6.8 Compatibility

Compatibility (CPA) is defined as the degree to which an innovation aligns with existing values, past experiences, and the needs of potential adopters (Rogers, 2003). According to IDT, compatibility is a key factor influencing technology adoption, as users are more likely to embrace innovations that fit seamlessly into their established practices. In the context of healthcare metaverse adoption, compatibility reflects how well the technology integrates with current teaching methods, clinical training, and learning needs.

A higher level of compatibility enhances PU because users are more inclined to adopt a system that complements their current educational and professional practices. If the healthcare metaverse aligns with medical students' learning styles and instructors' teaching methods, they will perceive it as a valuable tool for training and professional development. Similarly, compatibility positively influences PEOU, as users tend to find technologies more intuitive when they build on familiar experiences. Systems that require minimal adjustments to existing workflows reduce resistance and cognitive effort, making them easier to adopt. Additionally, compatibility directly affects BIU, as individuals are more likely to adopt innovations that seamlessly fit into their routines. A metaverse platform that aligns with current educational structures and skill development processes encourages adoption by students and faculty.

Empirical studies confirm the strong impact of compatibility on technology adoption. Research indicates that compatibility significantly influences PU, as users perceive systems that align with their needs as more beneficial (Al-Rahmi et al., 2019; Toraman & Geçit, 2023; Yap et al., 2024). Several studies further establish a direct positive link between compatibility and BIU, suggesting that alignment with existing practices enhances adoption likelihood (Ayanwale & Molefi, 2024; Lee et al., 2011; Raman et al., 2021). Additionally, research supports the influence of compatibility on PEOU, as users are more likely to find a system easy to use if it fits their prior experiences (Al-Rahmi et al., 2021; Al-Rahmi et al., 2019; Günay & Toraman, 2023).

H14: "Compatibility positively influences behavioral intention to use the healthcare metaverse in medical education."

H15: "Compatibility positively influences perceived usefulness of the healthcare metaverse in medical education."

H16: "Compatibility positively influences perceived ease of use of the healthcare metaverse in medical education."

### 2.6.9 Trialability

The intention to use technology is strongly connected to trialability (TRI), which refers to how effortless it is to try out and experiment with new technology. This includes factors such as the level of effort required and the risk involved in using the technology, as well as the ability to easily reverse or recover operations (Akour et al., 2022). IDT suggests that trialability enhances perceptions of an innovation's usefulness, as hands-on experience helps users assess its benefits. In the healthcare metaverse, trialability enables medical students and professionals to explore virtual simulations, engage with system functionalities, and determine its relevance to clinical education.

A higher degree of trialability is expected to enhance PU, as users who can experiment with the metaverse system firsthand are more likely to recognize its educational benefits. Engaging in a low-risk setting allows users to assess how the metaverse improves medical training, increasing their perception of its value.

Empirical research supports the role of trialability in technology adoption. Studies show that trialability enhances PU by providing users with direct experience, reducing uncertainty, and increasing confidence in the system's effectiveness (Van et al., 2022). Consequently, the following hypothesis could be formulated:

H17: "Trialability has a positive effect on perceived usefulness of healthcare metaverse in medical education."

### 2.6.10 Complexity

Complexity (CPL) is defined as "the degree to which the innovation is perceived as difficult to understand and use" (Rogers, 2003). According to IDT, the more complex a technology appears, the less likely users are to adopt it. Technologies perceived as overly difficult create cognitive barriers, increasing frustration and reducing engagement. In the healthcare metaverse, complexity reflects how challenging students and professors find the platform in terms of navigation, interface, and usability.

A higher level of complexity is expected to negatively impact PU, as users are less likely to see value in a system that is difficult to operate. When a technology is hard to navigate, users may struggle to recognize its benefits, leading to lower perceptions of its usefulness. Complexity also negatively affects PEOU, as users will find the metaverse less intuitive and more demanding. Increased effort in learning and operating the system discourages adoption. Consequently, complexity negatively influences BIU because users are less inclined to adopt a system they perceive as difficult and time-consuming.

Empirical research supports the negative effects of complexity on technology adoption. Studies confirm that higher complexity significantly reduces behavioral intention, as users avoid adopting systems they find difficult to use (Lee et al., 2011; Moorthy et al., 2019). Additionally, research shows a negative relationship between complexity and both PU and PEOU, indicating that as complexity increases, users perceive the system as less useful and harder to use (Lee et al., 2011; Parveen & Sulaiman, 2008). Therefore, drawing from the studies mentioned above, the following hypotheses can be proposed:

H18: "Complexity negatively influences behavioral intention to use the healthcare metaverse in medical education."

H19: "Complexity negatively influences perceived usefulness of the healthcare metaverse in medical education."

H20: "Complexity negatively influences perceived ease of use of the healthcare metaverse in medical education."

### 2.6.11 Technology readiness

Technology readiness (TR), a concept introduced by Parasuraman (2000), refers to "people's tendency to adopt and use new technologies to achieve their goals at home and at work". Achieving technology readiness can be challenging, as users often struggle to embrace new technologies. This readiness is influenced by positive drivers such as optimism and innovativeness, as well as negative factors like insecurity and discomfort (Almaiah, Alfaisal, Salloum, Al-Otaibi, et al., 2022). From this viewpoint, TR can be regarded as a holistic mental state shaped by various factors that either encourage or hinder an individual's inclination towards accepting new technologies (Lin et al., 2007).

The lack of a comprehensive understanding of TR underlines the need to consider TR in a broader context. A higher level of technology readiness positively influences PEOU, as individuals who are comfortable with new technologies require less effort to learn and navigate digital environments. Those with strong technological confidence tend to adapt more quickly, reducing perceived complexity and improving user experience. Additionally, technology readiness strengthens presence in virtual environments, as suggested by ESPT. Users who are more technologically prepared are better able to engage with immersive experiences, fostering a stronger sense of presence in the healthcare metaverse.

Almaiah, Alfaisal, Salloum, Al-Otaibi, et al. (2022) found that TR positively impacted the PEOU of digital information in education, indicating that users with higher technology readiness perceive systems as easier to use. Consequently, the following hypothesis could be formed:

H21: "Technology readiness has a positive effect on perceived ease of use of healthcare metaverse in medical education."

H22: "Technology readiness positively influences presence in the healthcare metaverse in medical education."

### 2.6.12 Imagination

The imagination (IM) is the ability to create mental images or concepts of things that do not physically exist, and it is fueled by the capacity of the mind to combine and build upon existing knowledge and new information (Almaiah, Alfaisal, Salloum, Hajjej, et al., 2022). In a VR setting, imagination refers to the human mind's ability to envision things that do not exist in the physical world (Huang et al., 2016). Using their imagination helps learners create mental images of new concepts and ideas that are not immediately apparent through their senses. By visualizing ideas creatively, learners can better understand what they are learning in metaverse environments and apply their knowledge to divergent and convergent thinking to build new understanding (Huang et al., 2010).

A higher level of imagination positively influences PU, as users who can mentally construct and interact with virtual medical scenarios are more likely to recognize the metaverse as a valuable learning tool. The ability to creatively engage with content enhances knowledge retention and skill acquisition, reinforcing the technology's PU. Additionally, imagination strengthens presence in virtual environments, a key concept in ESPT. The ability to vividly immerse oneself in a simulated environment enhances the sense of realism and interaction, making users feel more present within the metaverse.

Empirical research supports the role of imagination in technology adoption. Studies confirm that imagination significantly enhances PU by allowing users to creatively engage with VR-based learning content (Barrett et al., 2023; Huang et al., 2016). Furthermore, imagination has been found to strengthen presence by creating a deeper sense of immersion in virtual environments, which is critical for user engagement (Zhang et al., 2022). Consequently, the hypotheses could be formalized as follows:

H23: "Imagination has a positive effect on perceived usefulness of healthcare metaverse in medical education."

H24: "Imagination has a positive effect on presence of healthcare metaverse in medical education."

### 2.6.13 Technology anxiety

The term "technology anxiety" (TA) describes the uneasiness, dread and concern that individuals feel when utilizing technology or computer-based information systems. A related concept explored in the field of technology adoption is "computer anxiety," which is described as "the fear, apprehension, and hope people feel when considering use or actually using computer technology" (Scott and Rockwell, 1997). High levels of anxiety can make people less open to using information technologies (AlQudah et al., 2022). In the context of the healthcare metaverse, technology anxiety may arise due to unfamiliarity with immersive environments, complex interfaces, or concerns about errors in digital simulations.

A higher level of technology anxiety negatively influences PEOU because anxious users struggle with system navigation and interaction, perceiving the metaverse as more difficult than it actually is. Increased anxiety raises cognitive load, making it harder for users to develop confidence in their ability to use the platform efficiently. Additionally, technology anxiety negatively affects PU, as anxious individuals may be less likely to recognize the benefits of a system, focusing instead on its challenges or difficulties. This cognitive strain limits their ability to see the metaverse as an effective educational tool.

Empirical research supports the negative impact of technology anxiety on PEOU. Tung and Chang (2008) and Damar and Koksalmis (2023) indicate that higher anxiety levels reduce PEOU, as users feel more overwhelmed by digital interactions. Additionally, Almarzouqi et al. (2022a) discovered that technological anxiety negatively affects PU and PEOU. The hypotheses can be stated out as follows in light of the earlier research:

H25. "Technology anxiety negatively influences the perceived ease of use of healthcare metaverse in medical education."

H26. "Technology anxiety negatively influences the perceived usefulness of healthcare metaverse in medical education."

### 2.6.14 Moderating role of technology anxiety

The moderating role of technology anxiety in technology adoption has been relatively underexplored. While individuals may perceive a system as useful and easy to use, high levels of anxiety can still act as a barrier, reducing their willingness to adopt it. Users experiencing anxiety may hesitate to engage with a system, even if they recognize its benefits or find it easy to navigate.

As a result, technology anxiety weakens the relationship between PEOU and BIU, as anxious users may still avoid adoption despite finding the system intuitive. Similarly, technology anxiety moderates the effect of PU on BIU, as individuals who feel anxious about technology may resist adoption due to discomfort or fear, regardless of their perceived benefits. Empirical findings support these moderating effects, with Bhatt (2022) demonstrating that technology anxiety negatively influences the strength of these relationships in technology adoption contexts. Based on the findings of previous research, the hypotheses can be formulated as follows:

H27a. "Technology anxiety negatively moderates the relationship between the perceived usefulness and intention to use of healthcare metaverse in medical education."

H27b. "Technology anxiety negatively moderates the relationship between the perceived ease of use and intention to use of healthcare metaverse in medical education."

## 3 Methodology

## 3.1 Survey design

The questionnaire was divided into three components. The first section included a cover letter and an informed consent statement. The second section introduced participants to the concept of healthcare metaverse applications in medical education. Detailed information was provided, along with visual aids to help participants concretely visualize the potential uses of the healthcare metaverse. To further support understanding, two explanatory videos offered practical examples of how the healthcare metaverse can be used in medical education. This approach was crucial for participants who had no prior experience with these technologies, ensuring they could respond more knowledgeably to the survey questions. The final section collected demographic data and included questions related to the constructs examined in this research.

Ethical approval for the study was granted by the "Social Sciences and Human Research Ethics Committee of Istanbul Technical University" on March 30, 2023 (Project Number: 345). Prior to their participation, all participants gave their informed consent.

## 3.2 Participants

This study focuses on medical doctors and medical students enrolled in medical faculties in Turkey, who possess the necessary theoretical background and clinical orientation to assess the future integration of healthcare metaverse technologies. The healthcare metaverse is recognized as a game-changer in medical education, offering immersive simulations that enable students and professionals to practice essential clinical skills, including surgical procedures, in virtual environments (Rane et al., 2023). As future healthcare professionals, medical students are expected to increasingly engage with these digital and simulation-based learning tools. On the other hand, general practitioners are regarded not only as technology adopters but also as educators and decision-makers who influence organizational innovations. Including both populations offers a comprehensive perspective on adoption potential from both the trainee and educator viewpoints.

By focusing on participants from medical institutions in Turkey, this study captures perceptions within a national context where metaverse adoption in education remains in its early stages. Given the growing interest in technological readiness and immersive health education in Turkey, this sample provides valuable insights into the potential uptake of such technological tools within a transitioning health education ecosystem.

## 3.3 Data collection

Data were collected through an online survey administered via Google Forms between April 1 and June 10, 2023. A non-probability convenience sampling method was employed, prioritizing accessibility and voluntary participation. The survey link was distributed through the institutional mailing lists of medical schools in Turkey, targeting both current medical students and doctors. A total of 718 valid responses were collected, with no missing data or outliers. According to Hair et al. (2016), this sample size is sufficient for analysis. The sample consisted of 50.8% female respondents, with a minimum age of 18 years. A comprehensive overview of the demographic information is presented in Table 2.

Only 24.1% of participants reported having prior experience with metaverse applications in their medical education. This relatively low level of prior experience reflects the current state of adoption of healthcare metaverse technologies. Nevertheless, this finding provides valuable insight into the potential acceptance and readiness for future integration of these technologies into medical education. The primary objective of the study was to assess the intentions and perceptions of medical students and doctors concerning the use of healthcare metaverse in their education and professional practice.

**Table 2** Demographic structure of the sample

| *Age (years)* | | |
|---|---|---|
| Max: 61 | Min: 18 | |
| *Gender (%)* | | |
| Female: 50.8 | Male: 49.2 | |
| *Education level (%)* | | |
| School of Medicine 1st Year: 10.8 | School of Medicine 2nd Year: 16.3 | School of Medicine 3rd Year: 12.7 |
| School of Medicine 4th Year: 14.9 | School of Medicine 5th Year: 17.5 | School of Medicine 6th Year: 17.5 |
| Specialist: 16.6 | | |
| *Speciality (%)* | | |
| "Genetics": 8.2 | "Otorhinolaryngology": 16.4 | "Cytopathology and Embryology": 13.3 |
| "Medical Biochemistry": 19.1 | "Internal Medicine": 17.4 | "Basic Medical Sciences": 12.9 |
| "Breast Surgery": 8.4 | "Neurosurgery: 4.3 " | |
| *"Have you ever had experience using metaverse before? (%)":* | | |
| Yes: 24.1 | No: 75.9 | |
| *"Have you had previous experience of using the healthcare metaverse in medical education? (%)":* | | |
| Yes: 7.5 | No: 92.5 | |

### 3.4 Measures

A scale with 56 items was employed in the study, organized into nine constructs. Table 3 lists these constructs, along with their relevant items and references for each construct. Participants' opinions on each item were evaluated implementing a five-point Likert scale, where 1 corresponds to "strongly disagree," 2 to "disagree," 3 to "uncertain," 4 to "agree," and 5 to "strongly agree.". Since the model is literature-based and many elements in the questionnaire originate from previous research studies in English, the scale was originally designed in English. To guarantee precise translation, a double translation protocol (English–Turkish–English) was utilized (Hambleton, 1993), which engaged two native English speakers holding master's degrees in Turkish language. Following this, the questionnaire was subjected to pre-testing with 20 prospective participants to detect linguistic subtleties in each element, resulting in necessary adjustments to the phrasing. This pre-test also aimed to evaluate the comprehensibility of the survey and ensure the accuracy of the questions. Involving five medical doctors and two medical students with prior experience in healthcare metaverse applications, the pre-test revealed that these individuals fully grasped the instructions and questions presented in the survey. This finding reflects a positive outcome in terms of fulfilling the survey's objectives and meeting the participants' expectations.

**Table 3** Items associated with constructs

| Construct | Code | References | Items |
|---|---|---|---|
| Technology Readiness | TR1 | (Almaiah, Alfaisal, Salloum, Al-Otaibi, et al., 2022) | "I am ready to use metaverse in medical education for my information search and evaluation." |
| | TR2 | | "I am ready to accept metaverse in medical education if it makes accessing information easier." |
| | TR3 | | "I am ready to accept metaverse in medical education in order to integrate my theoretical knowledge with practical applications." |
| | TR4 | | "I am ready to use metaverse in medical education as it facilitates easier access to information." |
| Technology Anxiety | TA1 | (AlQudah et al., 2022; Tung & Chang, 2008) | "I have avoided healthcare metaverse because it is unfamiliar to me." |
| | TA2 | | "Using healthcare metaverse makes me feel uncomfortable." |
| | TA3 | | "Working with healthcare metaverse makes me anxious." |
| | TA4 | | "Healthcare metaverse are somewhat intimidating to me." |
| Learner-learner interaction | LL1 | (Kuo et al., 2014) | "I believe the healthcare metaverse would allow numerous interactions with fellow students about course content." |
| | LL2 | | "I expect to be able to communicate with fellow students about course content using communication tools (e.g., virtual spaces, avatars) in the healthcare metaverse." |
| | LL3 | | "I believe I would receive feedback from fellow students on course-related topics through communication features of the healthcare metaverse." |
| Learner-teacher interaction | LT1 | (Kuo et al., 2014) | "I expect that the healthcare metaverse would allow numerous interactions with the instructor about course-related topics." |
| | LT2 | | "I believe the healthcare metaverse would allow instructors to provide feedback through its communication tools when needed." |
| | LT3 | | "I expect that I would be able to ask my questions to the instructor using communication tools (e.g., virtual meetings, discussion boards) in the healthcare metaverse." |
| | LT4 | | "I expect instructors would respond to my |

questions in a timely manner using the communication features of the healthcare metaverse."

**Table 3** (continued)

| Construct | Code | References | Items |
|---|---|---|---|
| Relative advantage | RA1 | (Mikropoulos et al., 2022) | "Healthcare metaverse would be more advantageous in my future teaching than other technologies." |
| | RA2 | | "Healthcare metaverse would make my future teaching more effective than other technologies." |
| | RA3 | | "Healthcare metaverse is relatively efficient in my future teaching compared to existing technologies." |
| | RA4 | | "The use of healthcare metaverse offers new learning opportunities compared to existing technologies." |
| Trialability | TRI1 | (Jon et al., 2001; Martins et al., 2004; Rogers et al., 2014) | "I would like to try using healthcare metaverse before actual classes." |
| | TRI2 | | "It take time to get used to healthcare metaverse." |
| | TRI3 | | "I found healthcare metaverse useful after my trail." |
| | TRI4 | | "I am permitted to try healthcare metaverse for a long enough period." |
| | TRI5 | | "Before deciding to use healthcare metaverse, I would like to see its features first." |
| Complexity | CPL1 | (Laukkanen & Cruz, 2009; Tan & Teo, 2000) | "Using the healthcare metaverse requires a lot of mental effort." |
| | CPL2 | | "Using the healthcare metaverse requires technical skills." |
| | CPL3 | | "Using the healthcare metaverse can be frustrating." |
| Compatibility | CPA1 | (Chang & Tung, 2008) | "I think the healthcare metaverse is compatible with my studying purposes." |
| | CPA2 | | "I will use healthcare metaverse because it satisfies my expectations." |
| | CPA3 | | "I believe that the healthcare metaverse will süit my culture." |

| Construct | Code | References | Items |
|---|---|---|---|
| Imagination | IM1 | (Huang et al., 2016) | "I feel the healthcare metaverse helps me understand body structure better than a 2D system." |
| | IM2 | | "I feel the healthcare metaverse improves my understanding of the spatial relationships of body structure." |
| | IM3 | | "I feel the healthcare metaverse helps me better understand the relative positions of organs." |
| | IM4 | | "I feel the healthcare metaverse helps me better understand the real shape of organs." |

**Table 3** (continued)

| Construct | Code | References | Items |
|---|---|---|---|
| Presence | PRE1 | (Bailenson et al., 2004; Gao et al., 2017) | "There is a sense of human interaction in the metaverse." |
| | PRE2 | | "Socializing in the metaverse with a sense of reality." |
| | PRE3 | | "I can be aware of my presence in the metaverse. (Drop)" |
| | PRE4 | | "In the metaverse, the incarnation is sentient and alive to me." |
| Perceived Ease of Use | PEOU1 | (Davis, 1985; Davis et al., 1989; Doll et al., 1998; Venkatesh & Davis, 2000) | "I think healthcare metaverse is effortless." |
| | PEOU2 | | "I think I can use healthcare metaverse for different educational purposes since it's easy." |
| | PEOU3 | | "I think healthcare metaverse will be difficult to use in certain circumstances." |
| | PEOU4 | | "My interaction with healthcare metaverse would be clear and understandable." |
| | PEOU5 | | "It would be easy for me to become skillful at using healthcare metaverse applications." |
| Perceived Usefulness | PU1 | (Davis, 1985; Doll et al., 1998) | "I think healthcare metaverse is useful for live lectures and forums." |
| | PU2 | | "I think healthcare metaverse adds many advantages to my study." |
| | PU3 | | "Using healthcare metaverse would make it easier to accomplish my tasks." |
| | PU4 | | "Using healthcare metaverse would improve my productivity." |
| | PU5 | | "Using healthcare metaverse would increase my efficiency." |
| Satisfaction | SAT1 | (Donkor, 2011) | "I am satisfied with my learning from the healthcare metaverse environment." |
| | SAT2 | | "I find the healthcare metaverse to be effective in meeting the learning objectives in medical education." |

| | SAT3 | | "The healthcare metaverse has contributed greatly to my acquisition of relevant medical skills." |
| | SAT4 | | "The healthcare metaverse motivates me to spend more time studying and engaging with practical medical scenarios." |
| Behavioral Intention to use | BIU1 | (Al-Aulamie, 2013; Almarzouqi et al., 2022b) | "I will definitely use healthcare metaverse in my education." |
| | BIU2 | | "I intend to increase my use of healthcare metaverse in the future." |
| | BIU3 | | "I will use healthcare metaverse for limited educational purposes." |
| | BIU4 | | "For future studies I would use the healthcare metaverse." |

### 3.5 Data analysis

This research utilized the PLS-SEM methodology to examine the data that was obtained. PLS-SEM offers greater flexibility regarding data distribution and is applicable to a broad spectrum of sample sizes, ranging from small to large (Hair et al., 2011). This method has a wider range of applications as it does not require strict conditions such as normality of the data distribution. In addition, PLS-SEM is especially appropriate for evaluating intricate models that involve a large number of constructs and indicators (Hair et al., 2011). SmartPLS 4 software was utilized for the analysis. The structural model and the measurement model were both scrutinized as part of the two-stage evaluation technique used for the data assessment (Hair et al., 2017). In the initial stage, the measurement model was thoroughly analyzed. During this process, the reliability and validity of the selected constructs were rigorously evaluated. In the subsequent stage, the proposed relationships among the constructs were examined through the PLS-SEM method. This method allowed for the determination of the relationships between the constructs and the strength of these relationships.

## 4 Results

### 4.1 Measurement model

Several tests were performed to ascertain the accuracy and effectiveness of the measurement items and constructs. To evaluate construct validity, the item loadings were examined, with loadings above 0.6 considered desirable (Hair et al., 1998). All factor loadings exceeded 0.6. It is advised to set a minimum threshold of 0.7 for Cronbach's alpha and composite reliability (CR) to ensure the reliability and internal consistency of the constructs (Hair et al., 2016). Table 4 indicates that Cronbach's alpha and CR values for all constructs meet the threshold limit. Convergent validity was also evaluated by reviewing the "average variance extracted - AVE." It's generally accepted that an AVE threshold of 0.5 is appropriate (Hair et al., 2016). As indicated in Table 4, the AVE scores for each construct significantly exceeded the threshold of 0.5. This indicates that there is convergent validity among all constructs.

    Through a comparison of the square root of the average variance extracted (AVE) for each construct and the correlation between different constructs, discriminant validity was evaluated. The amount of variance that each construct captures is shown by the AVE. To establish sufficient

discriminant validity, it is expected that the diagonal elements (i.e., the square roots of AVE) should be higher than the off-diagonal elements (i.e., the correlations between constructs) (Fornell & Larcker, 1981). Table 5 provides an overview of the discriminant validity assessment. It shows that the diagonal elements, representing the square roots of AVE, exceed the off-diagonal elements, indicating that the constructs have satisfactory discriminant validity. This implies that each construct captures a significant amount of unique variance, distinct from other constructs in the measurement model. In addition to assessing the correlations and square roots of AVE, the "Heterotrait-Monotrait Ratio-HTMT" test was conducted to further examine discriminant validity. This test compares the correlation between constructs to a predetermined threshold, typically set at 0.85 (Henseler et al., 2015). The results of the HTMT test indicated that every single HTMT score was below the threshold (Table 6).

**Table 4** Reliability and validity

| Construct | Item code | Factor Loadings | Cronbach's alpha | CR | AVE |
|---|---|---|---|---|---|
| Behavioral intention to use | BIU1 | 0.929 | 0.836 | 0.891 | 0.673 |
| | BIU2 | 0.779 | | | |
| | BIU3 | 0.788 | | | |
| | BIU4 | 0.775 | | | |
| Satisfaction | SAT1 | 0.940 | 0.868 | 0.910 | 0.718 |
| | SAT2 | 0.817 | | | |
| | SAT3 | 0.812 | | | |
| | SAT4 | 0.814 | | | |
| Perceived usefulness | PU1 | 0.947 | 0.894 | 0.923 | 0.705 |
| | PU2 | 0.817 | | | |
| | PU3 | 0.815 | | | |
| | PU4 | 0.795 | | | |
| | PU5 | 0.815 | | | |
| Perceived ease of use | PEOU1 | 0.946 | 0.892 | 0.921 | 0.700 |
| | PEOU2 | 0.804 | | | |
| | PEOU3 | 0.780 | | | |
| | PEOU4 | 0.822 | | | |
| | PEOU5 | 0.822 | | | |
| Learner-learner interaction | LL1 | 0.942 | 0.859 | 0.914 | 0.780 |
| | LL2 | 0.855 | | | |
| | LL3 | 0.849 | | | |
| Learner-teacher interaction | LT1 | 0.949 | 0.889 | 0.924 | 0.752 |
| | LT2 | 0.838 | | | |
| | LT3 | 0.833 | | | |
| | LT4 | 0.843 | | | |
| Presence | PRE1 | 0.940 | 0.869 | 0.911 | 0.719 |
| | PRE2 | 0.813 | | | |
| | PRE3 | 0.821 | | | |
| | PRE4 | 0.812 | | | |
| Relative advantage | RA1 | 0.941 | 0.874 | 0.914 | 0.727 |
| | RA2 | 0.827 | | | |
| | RA3 | 0.819 | | | |
| | RA4 | 0.816 | | | |
| Compatibility | CPA1 | 0.944 | 0.847 | 0.907 | 0.766 |
| | CPA2 | 0.831 | | | |
| | CPA3 | 0.845 | | | |
| Trialability | TRI1 | 0.951 | 0.905 | 0.930 | 0.726 |
| | TRI2 | 0.827 | | | |
| | TRI3 | 0.825 | | | |

|  | TRI4 | 0.824 |  |  |  |
| --- | --- | --- | --- | --- | --- |
|  | TRI4 | 0.826 |  |  |  |
| Complexity | CPL1 | 0.927 | 0.759 | 0.859 | 0.673 |
|  | CPL2 | 0.782 |  |  |  |
|  | CPL3 | 0.740 |  |  |  |
| Technology readiness | TR1 | 0.948 | 0.881 | 0.919 | 0.739 |
|  | TR2 | 0.834 |  |  |  |
|  | TR3 | 0.828 |  |  |  |
|  | TR4 | 0.823 |  |  |  |
| Imagination | IM1 | 0.944 | 0.888 | 0.923 | 0.749 |
|  | IM2 | 0.842 |  |  |  |
|  | IM3 | 0.841 |  |  |  |
|  | IM4 | 0.831 |  |  |  |
| Technology anxiety | TA1 | 0.917 | 0.833 | 0.889 | 0.667 |
|  | TA2 | 0.777 |  |  |  |
|  | TA3 | 0.788 |  |  |  |
|  | TA4 | 0.777 |  |  |  |

**Table 5** Fornell-Larcker criterion results

| Construct | CPA | CPL | BIU | IM | LL | LT | PEOU | PRE | TRI | PU | RA | SAT | TA | TR |
|---|---|---|---|---|---|---|---|---|---|---|---|---|---|---|
| CPA | 0.875 | | | | | | | | | | | | | |
| CPL | -0.317 | 0.820 | | | | | | | | | | | | |
| BIU | 0.561 | -0.549 | 0.820 | | | | | | | | | | | |
| IM | 0.230 | -0.219 | 0.385 | 0.866 | | | | | | | | | | |
| LL | 0.271 | -0.287 | 0.409 | 0.241 | 0.883 | | | | | | | | | |
| LT | 0.298 | -0.223 | 0.409 | 0.188 | 0.463 | 0.867 | | | | | | | | |
| PEOU | 0.373 | -0.413 | 0.620 | 0.240 | 0.243 | 0.279 | 0.837 | | | | | | | |
| PRE | 0.417 | -0.383 | 0.618 | 0.294 | 0.324 | 0.284 | 0.398 | 0.848 | | | | | | |
| TRI | 0.279 | -0.251 | 0.420 | 0.515 | 0.231 | 0.215 | 0.258 | 0.276 | 0.852 | | | | | |
| PU | 0.395 | -0.346 | 0.573 | 0.674 | 0.334 | 0.278 | 0.380 | 0.384 | 0.703 | 0.840 | | | | |
| RA | 0.436 | -0.387 | 0.576 | 0.278 | 0.259 | 0.320 | 0.410 | 0.375 | 0.261 | 0.415 | 0.853 | | | |
| SAT | 0.391 | -0.401 | 0.589 | 0.294 | 0.656 | 0.667 | 0.410 | 0.391 | 0.325 | 0.422 | 0.407 | 0.848 | | |
| TA | -0.217 | 0.270 | -0.417 | -0.175 | -0.187 | -0.195 | -0.660 | -0.266 | -0.190 | -0.255 | -0.288 | -0.304 | 0.817 | |
| TR | 0.250 | -0.349 | 0.441 | 0.216 | 0.182 | 0.194 | 0.710 | 0.273 | 0.191 | 0.291 | 0.303 | 0.280 | -0.514 | 0.860 |

**Table 6** Heterotrait-monotrait ratios (HTMT ratios)

| Construct | CPA | CPL | BIU | IM | LL | LT | PEOU | PRE | TRI | PU | RA | SAT | TA | TR | TA x PU | TA x PEOU |
|---|---|---|---|---|---|---|---|---|---|---|---|---|---|---|---|---|
| CPA | | | | | | | | | | | | | | | | |
| CPL | 0.362 | | | | | | | | | | | | | | | |
| BIU | 0.635 | 0.639 | | | | | | | | | | | | | | |
| IM | 0.253 | 0.254 | 0.430 | | | | | | | | | | | | | |
| LL | 0.305 | 0.318 | 0.460 | 0.267 | | | | | | | | | | | | |
| LT | 0.337 | 0.249 | 0.458 | 0.209 | 0.514 | | | | | | | | | | | |
| PEOU | 0.416 | 0.473 | 0.700 | 0.262 | 0.266 | 0.307 | | | | | | | | | | |
| PRE | 0.469 | 0.446 | 0.707 | 0.324 | 0.365 | 0.318 | 0.441 | | | | | | | | | |
| TRI | 0.307 | 0.291 | 0.469 | 0.564 | 0.252 | 0.235 | 0.281 | 0.300 | | | | | | | | |
| PU | 0.440 | 0.395 | 0.643 | 0.742 | 0.370 | 0.307 | 0.417 | 0.424 | 0.770 | | | | | | | |
| RA | 0.495 | 0.447 | 0.648 | 0.308 | 0.286 | 0.355 | 0.453 | 0.419 | 0.287 | 0.462 | | | | | | |
| SAT | 0.442 | 0.462 | 0.669 | 0.328 | 0.738 | 0.747 | 0.457 | 0.439 | 0.360 | 0.468 | 0.456 | | | | | |
| TA | 0.244 | 0.317 | 0.483 | 0.197 | 0.212 | 0.218 | 0.745 | 0.302 | 0.217 | 0.287 | 0.325 | 0.349 | | | | |
| TR | 0.280 | 0.413 | 0.498 | 0.242 | 0.202 | 0.217 | 0.787 | 0.302 | 0.211 | 0.320 | 0.337 | 0.312 | 0.584 | | | |
| TA x PU | 0.047 | 0.111 | 0.097 | 0.137 | 0.032 | 0.034 | 0.187 | 0.045 | 0.141 | 0.193 | 0.061 | 0.110 | 0.267 | 0.127 | | |
| TA x PEOU | 0.156 | 0.257 | 0.299 | 0.144 | 0.128 | 0.164 | 0.502 | 0.190 | 0.089 | 0.170 | 0.234 | 0.200 | 0.616 | 0.402 | 0.383 | |

## 4.2 Hypotheses

The validity of the hypotheses was evaluated using the PLS-SEM approach. The main hypotheses, associated t-statistics and standardized parameters are shown in Table 7. This method was utilized to identify the significance and direction of the relationships in the study.

The findings address RQ4, which investigates the influence of PU and PEOU on BIU. The results show that PU significantly affects BIU ($\beta = 0.169$, $p < 0.05$; H2 supported), demonstrating that users are more likely to adopt metaverse technologies when they find them beneficial. PU also significantly influences satisfaction ($\beta = 0.108$, $p < 0.05$; H3 supported), reinforcing the idea that usefulness enhances overall user satisfaction. Moreover, PEOU significantly predicts BIU ($\beta = 0.226$, $p < 0.05$; H4 supported), SAT ($\beta = 0.142$, $p < 0.05$; H5 supported), PU ($\beta = 0.080$, $p < 0.05$; H6 supported), indicating that a user-friendly metaverse environment improves adoption likelihood, perceived benefits, and satisfaction.

RQ3 is examined through the impact of LTI and LLI on satisfaction and its subsequent effect on behavioral intention. The findings also reveal that LTI and LLI positively impact satisfaction ($\beta = 0.405$, $p < 0.05$; H7 supported; $\beta = 0.381$, $p < 0.05$; H8 supported), emphasizing the role of social engagement in shaping positive user experiences. Furthermore, the results indicate that higher satisfaction significantly increases BIU ($\beta = 0.164$, $p < 0.05$; H1 supported), suggesting that users who experience greater satisfaction in metaverse-based learning environments are more likely to adopt the technology. These findings highlight the importance of fostering meaningful LTI and LLI as an indirect pathway to enhancing adoption intention through improved satisfaction.

Addressing RQ2, the results demonstrate that perceived presence significantly influences BIU ($\beta = 0.230$, $p < 0.05$; H9 supported), but its effect on SAT is marginally non-significant ($\beta = 0.055$, $p = 0.055$; H10 not supported). While the result does not meet the conventional significance threshold of $p < 0.05$, the p-value of 0.055 suggests a potential trend toward significance, meaning that presence may still contribute to satisfaction in some cases. This result may indicate that with a larger sample size, a significant effect might be observed. Future studies could explore this relationship further by incorporating additional moderating variables or refining the measurement of perceived presence. Furthermore, RQ2 is extended by examining the role of imagination in influencing PU and PRE. The results show that imagination significantly influences PU ($\beta = 0.374$, $p < 0.05$; H23 supported) and PRE ($\beta = 0.247$, $p < 0.05$; H24 supported), reinforcing the role of imagination in shaping user perceptions of usefulness and immersive presence.

RQ1 is examined through the innovation attributes derived from IDT. The findings show that RA, CPA, CPL, and TRI significantly influence adoption factors. Specifically, RA significantly impacts BIU ($\beta = 0.142$, $p < 0.05$; H11 supported), PU ($\beta = 0.104$, $p < 0.05$; H12 supported), and PEOU ($\beta = 0.087$, $p < 0.05$; H13 supported), indicating that the perceived benefits of the metaverse influence both its PEOU and PU, ultimately driving adoption. Similarly, CPA significantly influences BIU ($\beta = 0.142$, $p < 0.05$; H14 supported), PU ($\beta = 0.099$, $p < 0.05$; H15 supported), and PEOU ($\beta = 0.119$, $p < 0.05$; H16 supported), highlighting that individuals who perceive metaverse technologies as compatible with their learning styles and needs are more likely to adopt them, perceive them as useful, and find them easy to use. Additionally, trialability positively affects PU ($\beta = 0.421$, $p < 0.05$; H17 supported), suggesting that users who can experiment with the technology are more likely to perceive it as beneficial. In contrast, complexity negatively influences BIU ($\beta = -0.145$, $p < 0.05$; H18 supported), PU ($\beta = -0.054$, $p <$ 0.05; H19 supported), and PEOU (β = -0.092, p < 0.05; H20 supported), indicating that higher complexity reduces adoption intention, PU, and PEOU perceptions. These findings suggest that simplified, user-friendly metaverse interfaces may enhance adoption rates and improve user experiences.

RQ5, which examines the role of technology readiness in shaping PEOU and presence, is also addressed in this section. The results confirm that TR significantly affects both PEOU (β = 0.437, p < 0.05: H21 supported) and PRE (β = 0.220, p < 0.05; H22 supported), indicating that users who are more technologically prepared not only find metaverse technologies easier to use but also experience a stronger sense of presence within the virtual environment.

Regarding technology-related anxieties, RQ6 is examined here by investigating direct effects of technology anxiety. The findings show that technology anxiety significantly reduces PEOU (β = -0.359, p < 0.05; H25 supported), indicating that users who experience higher levels of anxiety find the metaverse environment less intuitive and harder to use. However, technology anxiety does not significantly influence PU (β = 0.016, p > 0.05; H26 not supported), suggesting that while anxiety impacts ease of use perceptions, it does not alter users' perceptions of usefulness.

Figure 2 shows the path coefficients and R-Square scores. The R-Square ($R^2$) value for BIU is 0.718, indicating that 71.8% of the variance in BIU is explained by the proposed model, demonstrating strong explanatory power. The results confirm that the integration of multiple theoretical perspectives provides deeper insights into metaverse adoption in medical education, surpassing the predictive capabilities of traditional technology acceptance models.

**Table 7** Hypothesis tests results

| Hypothesis | Path | Path coefficient (β coefficient) | *T* statistics | Decision |
| --- | --- | --- | --- | --- |
| H1 | SAT -> BIU | 0.164* | 6.048 | "Supported" |
| H2 | PU -> BIU | 0.169* | 6.547 | "Supported" |
| H3 | PU -> SAT | 0.108* | 4.010 | "Supported" |
| H4 | PEOU -> BIU | 0.226* | 7.363 | "Supported" |
| H5 | PEOU -> SAT | 0.142* | 5.384 | "Supported" |
| H6 | PEOU -> PU | 0.080* | 3.046 | "Supported" |
| H7 | LTI -> SAT | 0.405* | 16.097 | "Supported" |
| H8 | LLI -> SAT | 0.381* | 15.382 | "Supported" |
| H9 | PRE -> BIU | 0.230* | 8.785 | "Supported" |
| H10 | PRE -> SAT | 0.055 | 1.915 | "Not supported" |
| H11 | RA -> BIU | 0.142* | 5.372 | "Supported" |
| H12 | RA -> PU | 0.104* | 3.756 | "Supported" |
| H13 | RA -> PEOU | 0.087* | 2.945 | "Supported" |
| H14 | CPA -> BIU | 0.142* | 5.689 | "Supported" |
| H15 | CPA -> PU | 0.099* | 3.724 | "Supported" |
| H16 | CPA -> PEOU | 0.119* | 4.600 | "Supported" |
| H17 | TRI -> PU | 0.421* | 17.368 | "Supported" |

| Hypothesis | Path | Path coefficient (β coefficient) | T statistics | Decision |
|---|---|---|---|---|
| H18 | CPL -> BIU | -0.145* | 5.617 | "Supported" |
| H19 | CPL -> PU | -0.054* | 2.209 | "Supported" |

**Table 7** (continued)

| Hypothesis | Path | Path coefficient (β coefficient) | T statistics | Decision |
|---|---|---|---|---|
| H20 | CPL -> PEOU | -0.092* | 3.568 | "Supported" |
| H21 | TR -> PEOU | 0.437* | 17.741 | "Supported" |
| H22 | TR -> PRE | 0.220* | 6.154 | "Supported" |
| H23 | IM -> PU | 0.374* | 15.852 | "Supported" |
| H24 | IM -> PRE | 0.247* | 6.378 | "Supported" |
| H25 | TA -> PEOU | -0.359* | 15.057 | "Supported" |
| H26 | TA -> PU | -0.016 | 0.518 | "Not supported" |

*: $p < 0.05$

## 4.3 Moderation effect analysis

To address RQ6, which examines how technology anxiety moderates key adoption relationships, the moderating effects of technology anxiety on PU and BIU (H27a) and PEOU and BIU (H27b) were tested (Table 8). The results indicate that neither moderation effect was significant (β = -0.038, p > 0.05 for H27a; β = -0.010, p > 0.05 for H27b). These findings suggests that while anxiety influences ease of use perceptions, it does not significantly alter the effect of PU or PEOU on adoption intention. These findings imply that while anxiety-related concerns may shape usability perceptions, they do not directly moderate the likelihood of adoption, warranting further investigation into how anxiety operates in metaverse learning environments.

**Table 8** Moderation effect analysis results

| Hypothesis | Path | Path coefficient (β coefficient) | T statistics | Decision |
|---|---|---|---|---|
| H27a | TA*PU -> BIU | -0.038 | 1.715 | "Not supported" |
| H27b | TA*PEOU -> BIU | -0.010 | 0.424 | "Not supported" |

## 4.4 Mediation effect analysis

In mediation, an intermediate variable, known as the mediator, is considered to help explain how or why an independent variable affects an outcome (Gunzler et al., 2013).

RQ7, which explores whether PEOU mediates the relationship between technology anxiety and PU, is examined in this section. According to the hypothesis test results, the effect of technology anxiety on PU was not found to be significant. Therefore, based on the relationships within the model, an analysis was conducted under the assumption that technology anxiety may influence PU through PEOU, which serves as a mediator variable. To examine this relationship, a sub-model was established between technology anxiety and PU.

As shown in Table 9, the relationship between TA and PU is significant (p < 0.05). However, when both TA and PEOU are regressed on PU, the relationship between TA and PU becomes insignificant (p > 0.05). This confirms that PEOU fully mediates the effect of technology anxiety on PU, meaning that users with higher technology anxiety perceive metaverse environments as less intuitive, which in turn reduces their perception of usefulness.

**Table 9** Mediation effect analysis results

| Mediation effect analysis | Relationship | Standardized path coefficient (β coefficient) |
| --- | --- | --- |
| Sub-model linking technology anxiety to PU | "Technology anxiety -> Perceived usefulness" | -0.263* |
| Analysis of the mediating role of PEOU | "Technology anxiety -> Perceived usefulness" | -0.009 |
|  | "Technology anxiety -> Perceived ease of use" | -0.659* |
|  | "Perceived ease of use -> Perceived usefulness" | 0.375* |

*: p < 0.05*

## 5  Discussion and conclusions

This study investigated the factors influencing the acceptance of healthcare metaverse technologies in medical education by integrating multiple theoretical perspectives. Unlike prior research that primarily relied on TAM and UTAUT2, this study expanded the theoretical foundation by incorporating IDT, ESPT, and the EQuiv. Using PLS-SEM, the study validated a model explaining 71.8% of the variance in behavioral intention, demonstrating strong explanatory power.

The findings confirm core TAM constructs, such as PU, PEOU, and satisfaction, as significant predictors of users' behavioral intentions toward metaverse adoption. This study further extends previous research by highlighting the critical roles of relative advantage, compatibility, trialability, and perceived presence, which are derived from IDT and ESPT. These factors substantially influence users' perceptions and acceptance of metaverse-based educational platforms. Additionally, learner-instructor and learner-learner interactions, based on EQuiv, significantly impact user satisfaction, emphasizing the value of social connectivity and engagement in virtual learning environments.

The study also identified perceived presence as a key predictor of adoption. While presence significantly influenced behavioral intention, its effect on satisfaction was marginally non-significant (p = 0.055). This suggests that while immersion in a virtual environment enhances adoption, its contribution to overall user satisfaction may depend on other contextual factors such as interface design, system reliability, and individual cognitive load. Future studies could explore how different levels of presence influence learning effectiveness and user engagement over time.

An essential contribution of this study is the detailed analysis of technology anxiety and its indirect role in adoption decisions. Findings indicate that technology anxiety significantly reduces PEOU but does not directly impact PU. Moreover, moderation effects of TA on relationships between PU, PEOU, and behavioral intention were not supported, suggesting that while anxiety influences usability perceptions, it does not directly moderate the intention to adopt. The mediation analysis further demonstrated that PEOU fully mediates the relationship between technology anxiety and PU, indicating that improving system usability can effectively mitigate anxiety-related barriers.

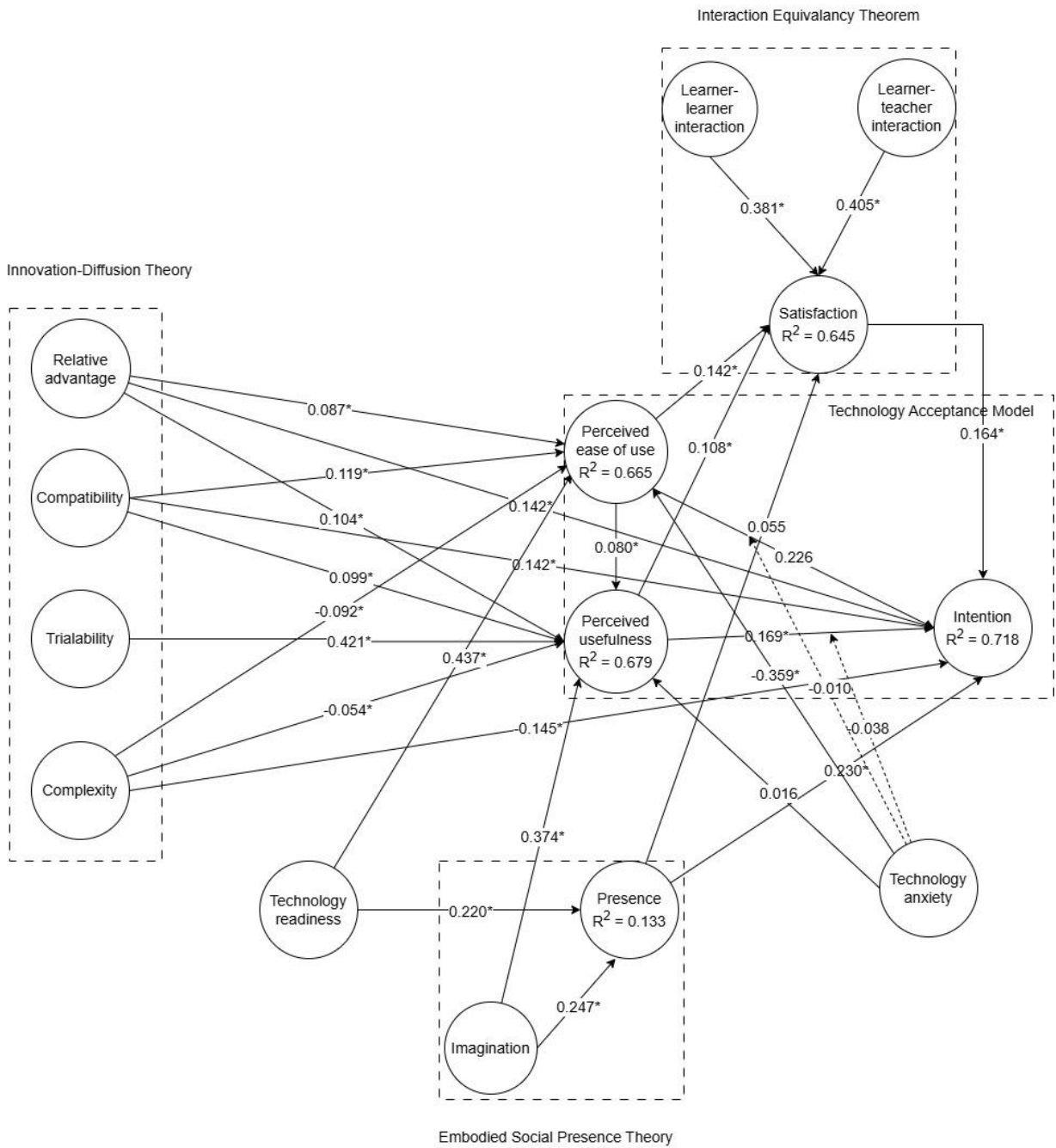

**Fig. 2** Path coefficients in the research model with statistical significance (*p<0.05)

These results highlight fundamental differences between metaverse technologies and traditional digital learning tools, emphasizing the necessity of a more sophisticated adoption model tailored to the unique attributes of metaverse environments. The integration of IDT, ESPT, and EQuiv expands theoretical understanding and provides deeper insights into the complexities of metaverse adoption in medical education.

To sum up, all the research questions formulated in this study were systematically addressed throughout the paper. RQ1 to RQ5 were examined in the 4.2 Hypotheses subsection, where the direct effects of key factors such as PU, PEOU, satisfaction, innovation attributes, and technology readiness on behavioral intention were validated. The role of technology anxiety in the adoption process, both in its direct effects and its moderating role, was analyzed in subsection 4.4 and further explored in 4.3 Moderation effect analysis subsection, addressing RQ6. Furthermore, RQ7 was examined in 4.4 Mediation effect analysis subsection, where the impact of satisfaction on behavioral intention was explored, emphasizing how social engagement enhances adoption through improved user experience. By resolving these research questions, this study provides a comprehensive understanding of metaverse adoption in medical education, offering key insights that contribute to both theoretical advancements and practical implementation strategies.

## 5.1 Theoretical implications

This study offers an enriched theoretical framework for understanding the adoption of metaverse technologies in medical education by integrating several complementary theoretical perspectives: IDT, ESPT, EQuiv, TAM, technology readiness, and technology anxiety. The synthesized model provides more comprehensive insights compared to traditional technology acceptance models by capturing the unique characteristics of the metaverse environment.

The inclusion of compatibility and relative advantage from IDT provides deeper insights into how healthcare professionals perceive metaverse technologies in relation to their existing professional practices. These findings highlight the necessity of aligning technological innovation with established clinical and educational workflows to encourage adoption. Furthermore, the significant impacts of trialability and complexity underline the importance of intuitive and accessible designs, emphasizing that ease of experimentation and minimal cognitive load significantly improve technology adoption.

Integrating ESPT constructs, specifically embodied presence and imagination, extends theoretical understanding by highlighting the experiential dimensions of the metaverse. The significant influence of presence on adoption intention underscores the importance of immersive experiences, while imagination's positive effect on PU and presence reveals the pivotal role of users' cognitive engagement and creativity in driving adoption.

Incorporating TR and TA extends existing theoretical perspectives, especially concerning individual user differences. The significant negative relationship between technology anxiety and PEOU reinforces the nuanced understanding that anxiety primarily affects usability perceptions rather than perceived benefits. The mediation analysis further illustrates how PEOU mediates the influence of anxiety on PU, providing a clearer theoretical contribution regarding how anxiety specifically impacts adoption processes in metaverse contexts.

Additionally, the Interaction Equivalency Theorem highlights learner-teacher and learner-learner interactions as critical factors shaping user satisfaction. This reinforces the theoretical importance of facilitating robust social interactions in virtual educational contexts to ensure sustained user engagement.

## 5.2 Practical implications

From a practical standpoint, the findings offer significant insights for healthcare educators, institutions, and technology developers. Firstly, ensuring compatibility with healthcare workflows and clearly demonstrating relative advantages over conventional training methods should become a primary consideration for implementation strategies. Institutions should actively promote trialability through pilot programs and training sessions, reducing perceived complexity by providing intuitive interfaces and clear guidance to enhance technology acceptance.

The critical role of embodied presence and imagination in influencing user perceptions suggests practical implications for designers and educators. Metaverse platforms should optimize immersive features that foster users' sense of presence, integrating creative visualization techniques into educational modules to enhance cognitive engagement and interactive learning experiences.

Addressing technology anxiety should also be a priority for stakeholders. Our study highlights that learner-learner and learner-teacher interactions serve as natural peer support mechanisms, helping users navigate new technological environments with greater confidence. These interactions provide structured opportunities for students and professionals to exchange experiences, ask questions, and learn from each other, thereby reducing uncertainty and anxiety. Institutions should implement peer mentoring programs, where experienced users guide newcomers through collaborative learning activities, such as team-based simulations, peer-led training sessions, and instructor-moderated discussion forums. Additionally, structured stepwise exposure to metaverse tools, starting with low-pressure training environments before full integration, can significantly enhance user readiness and reduce apprehension.

Furthermore, fostering technology readiness through preparatory workshops can substantially improve user perceptions and encourage widespread adoption of metaverse technologies. Finally, enhancing interactions between learners and teachers, as well as among peers, should be a key design principle. Platforms must prioritize interactive capabilities to create dynamic, engaging, and socially supportive virtual learning environments, directly translating to higher satisfaction and sustained user engagement.

## 5.3 Limitations and future research

Despite its significant contributions, this study has several limitations. Firstly, the cross-sectional nature of the research limits causal inferences. A longitudinal approach could provide deeper insights into how perceptions and adoption intentions evolve over time. Secondly, the marginally non-significant relationship between perceived presence and satisfaction indicates potential sample size limitations. Future studies could address this by employing larger and more diverse participant groups.

Additionally, this study focused primarily on healthcare education contexts, limiting generalizability. Exploring other medical and clinical contexts or comparative studies across different healthcare disciplines could enhance the external validity of the findings.

Future research should further explore the nuanced role of technology anxiety by considering additional moderating or mediating variables. Longitudinal studies examining temporal changes in adoption behavior would significantly contribute to understanding sustained engagement with metaverse technologies.

Moreover, qualitative methods such as interviews or case studies could provide richer insights into specific challenges and facilitators identified through quantitative measures. Additionally, comparative studies across different healthcare disciplines or varying educational levels could validate the robustness and applicability of the proposed theoretical framework.

Finally, future studies could investigate more deeply the dimensions of presence and imagination, exploring specific platform features or educational interventions that effectively enhance these constructs, thus offering practical guidance for optimizing metaverse-based medical education.